\documentclass[twocolumn]{aastex63}

\newcommand{\forse}{\textsc{ForSE }}
\newcommand{\forsee}{\textsc{ForSE}}

\usepackage{scalerel}

\shorttitle{\forsee: Foreground Scale Extender}
\shortauthors{Krachmalnicoff, N. \& Puglisi, G.}

\begin{document}

\title{ForSE: a GAN based algorithm for extending CMB foreground models to sub-degree angular scales}

\correspondingauthor{Nicoletta Krachmalnicoff}
\email{nkrach@sissa.it}

\author{Nicoletta Krachmalnicoff}
\affiliation{SISSA, Via Bonomea 265, 34136 Trieste, Italy}
\affiliation{INFN, Via Valerio 2, 34127 Trieste, Italy}
\affiliation{IFPU, Via Beirut 2, 34014 Trieste, Italy}

\author{Giuseppe Puglisi}
\affiliation{Space Sciences Laboratory,  University of California, Berkeley, CA 94720,  USA}
\affiliation{Computational Cosmology Center, Lawrence Berkeley National Laboratory, Berkeley, CA 94720, USA}

\begin{abstract}

We present \forse (Foreground Scale Extender), a novel Python package which aims at overcoming the current limitations in the simulation of diffuse Galactic radiation, in the context of Cosmic Microwave Background experiments (CMB). \forse exploits the ability of generative adversarial neural networks (GANs) to learn and reproduce complex features present in a set of images, with the goal of simulating realistic and non-Gaussian foreground radiation at sub-degree angular scales. This is of great importance in order to estimate the foreground contamination to lensing reconstruction, de-lensing and primordial $B$-modes, for future CMB experiments. We applied this algorithm to Galactic thermal dust emission in both total intensity and polarization. Our results show how \forse is able to generate small scale features (at 12 arc-minutes) having as input the large scale ones (80 arc-minutes). The injected structures have statistical properties, evaluated by means of the Minkowski functionals, in good agreement with those of the real sky and which show the correct amplitude scaling as a function of the angular dimension. The obtained thermal dust Stokes $Q$ and $U$ full sky maps as well as the \forse package are publicly available for download.
\end{abstract}


\section{Introduction}
\label{sec:intro}

In the last $\sim$ 50 years, observations of the Cosmic Microwave Background (CMB)  have allowed to unveil the physics and the evolution history of our Universe with an accuracy that was unimaginable when the CMB radiation was first detected in 1965 \citep{PW65}. Thanks to the measurements of three dedicated satellites and of several ground based and ballon-borne experiments that have successfully taken data over the decades, the CMB total intensity anisotropies have been constrained with great precision. The accuracy achieved on the main cosmological parameters is today at the level of percent or sub-percent, returning a picture of our Universe in large agreement with the $\Lambda CDM$ (Lambda cold dark matter) cosmological model \citep{2020A&A...641A...6P, 2020arXiv200707288A}.\par

 After the release, in 2018, of the legacy data of the Planck satellite \citep{2020A&A...641A...1P}, CMB observations have officially entered a new era. The goal of the current dedicated ground based experiments, as well as of the future satellites, is to measure with high accuracy the anisotropies in the CMB polarized signal, and to constrain its power spectra at all the angular scales, from tens of degrees to few arc-minutes.\par
 
CMB is partially linear polarized due to scattering events of photons against free electrons in the primordial Universe and in  presence of quadrupole anisotropies. The resulting polarization field, observed through the Stokes $Q$ and $U$ parameters, can be decomposed into a gradient ($E$-mode) and a curl ($B$-mode) component, respectively even and odd under parity inversion  \citep{1997PhRvD..55.1830Z, 1997PhRvD..55.7368K}. $E$-modes have been measured by several experiments in the past years, and their power spectrum as well as their correlation with the total intensity ($T$) field have been reconstructed up to the sub-degree angular scales \citep{2020A&A...641A...5P}. On the other hand, it  is in the $B$-mode signal that potential breakthrough discoveries might be hidden. \par
 
Primordial $B$-modes are sourced by gravitational (tensor) perturbations in the early Universe, contrary to density (scalar) perturbations that can only produce the $E$-mode polarization pattern. The presence of a background of primordial gravitational waves (GWs), as a manifestation of quantum gravity, is predicted by the Inflationary paradigms, i.e. a  family of theories that hypothesize a phase of accelerated expansion in the very early Universe \citep{1981PhRvD..23..347G, 1982PhLB..108..389L, 1979JETPL..30..682S}. The detection of the imprint of these GWs in the curl component of the CMB polarized signal would not only be fundamental to shed light on Inflation, but would also have a crucial impact on several fields in fundamental Physics. Moreover, more exotic phenomena such as cosmic strings and other topological defects in the primordial universe could also have left a trace into the CMB $B$-mode signal \citep{2003PhLB..563....6J, 2006PhRvD..74f3523S, 2015SchpJ..1031682V}.\par

The primordial $B$-mode power spectrum is characterized by the presence of two main bumps at intermediate and large angular scales: (i) the recombination bump, at multipoles $\ell\sim80$ (corresponding to an angular dimension of about $2^{\circ}$ in the sky); (ii) the reionization bump, located at $\ell\lesssim20$ (tens of degrees in the sky) and whose presence is due to a second phase of scattering events that CMB photons undertook during the reionization era.  The amplitude of the primordial spectrum is parametrized by the tensor-to-scalar ratio $r$, whose value varies significantly among different inflationary models \citep{2009arXiv0907.5424B}. The current upper limit is $r<0.06$ (95\% confidence level) coming from the combination of the BICEP2/Keck Array data with foreground observations from Planck and WMAP \citep{2018PhRvL.121v1301B}; this limit can be further shrunk to $r<0.044$ if full sky CMB data from Planck are also considered \citep{2020arXiv201001139T} \par

At smaller angular scales, the $B$-mode polarization pattern is generated as a consequence of gravitational lensing. As a matter of fact, CMB photons get deflected along their path, due to the presence of gravitational potentials generated by forming cosmological structures in the Universe. This effect causes a leak of the existing $E$-modes (generated by the primordial scalar perturbations) into the $B$ ones. The resulting lensing $B$-mode power spectrum shows a broad peak, that has its maximum at $\ell\sim1000$, corresponding to an angular scale of few arc-minutes. The amplitude of the lensing spectrum at $\ell\simeq100$ is equivalent to the one of a primordial $B$-mode signal with $r\simeq0.02$, and thus needs to be taken into account when targeting smaller values of $r$. The $B$-mode lensing signal has been first detected by the South Pole Telescope (SPTpol) in cross correlation with Cosmic Infrared Background (CIB) data \citep{2013PhRvL.111n1301H}, and subsequently the lensing peak has been successfully measured by the PolarBear experiment \citep{PolarBear14}, the Atacama Cosmology Telescope \citep[ACTpol]{2017JCAP...06..031L}, and others. \par

In the recent years, several experiments have been built or are currently being designed with the primary goal of observing the primordial $B$-modes and constrain the value of the $r$ parameter. Ground based experiments are focusing on the observation of the CMB polarization at intermediate and small angular scales, in order to detect the $B$-modes recombination bump. Among these experiment the Simons Observatory (SO) is currently being deployed with observations that will begin $\sim2022$ from the Atacama desert \citep{2019JCAP...02..056A}. SO will consist in three small-aperture 0.5-m telescopes (SATs) and one large-aperture 6-m telescope (LAT). The SATs will target the degree angular scales (corresponding to multipoles $30\lesssim\ell\lesssim500$) in about 10\% of the sky, while LAT observations will focus in the arc-minute angular scales ($\ell\gtrsim1000$) covering about 40\% of the sky. The combination of the measurements of the two instruments is necessary to achieve the goal of constraining the tensor-to-scalar ratio: LAT observations at small scales will allow to reconstruct the lensing potential that causes the CMB lensing distortion, and consequently will permit to perform de-lensing of the SAT maps. With this procedure SO will reach the target sensitivity on $r$ with $\sigma(r)\simeq0.003$. Similarly, but on a longer time scale (2027), the Stage-IV network of ground-based observatories (CMB-S4) will combine observations at intermediate and small angular scales to target detection of primordial $B$-modes with $r\gtrsim0.001$ \citep{2016arXiv161002743A}. A similar target on the tensor-to-scalar ratio is forecasted for the LiteBIRD satellite,  which has been selected as strategic large mission by the JAXA (Japan Aerospace Exploration Agency) space agency and will observe the CMB polarized signal at very large angular scales, with full sky coverage from the space ($\ell\lesssim200$), with operation starting in $\sim2028$ \citep{2020JLTP..199.1107S}. \par

One of the greatest challenges in the detection of primordial $B$-modes is represented by the contamination coming from Galactic foregrounds. In fact, polarized radiation from synchrotron and thermal dust emission pollutes the CMB tensor modes at the degree scales, everywhere in the sky and at every possible observation frequency \citep{2003NewAR..47.1127B, 2007ApJS..170..335P, 2018A&A...618A.166K, 2016A&A...588A..65K, PIP-XXX}. Several algorithm that perform the so-called \emph{component separation}, with the goal of getting a clean CMB maps have been developed in the recent years, including blind or parametric methods, in pixel or harmonic domain \citep[see for example:][]{planck2015-IX, 2016PhRvD..94h3526S}.\par

At smaller angular scales the Galactic emission amplitude diminishes but it is still strong enough to significantly contaminate the CMB lensing signal. This must be taken into account as the effect of foreground on lensing reconstruction and delensing could lead to residual spurious signals preventing the detection of primordial $B$-modes \citep{2020JCAP...06..030B}.\par

The properties of both thermal dust and synchrotron emission are known with reasonable accuracy at large angular scales ($\gtrsim1^{\circ}$) thanks to the full sky multi-frequency observations of the WMAP and Planck satellites, and these data are used to build our large scale foreground models and templates \citep{planck2015-X}. On the other hand, we do not have data with the sensitivity required to characterize polarized foregrounds at sub-degree resolution on large portion of the sky. Nonetheless, we know that Galactic emission have a non trivial statistics, with a non-Gaussian, non-stationary signal and a complicated morphology that reflects the complexity of the Galactic magnetic field \citep{2019JCAP...10..056C}; we expect this to be true at all angular scales.\par

 Being able to simulate the complexity of foregrounds, from the larger scales to the arc-minute ones, is fundamental, in order to understand their impact on both lensing and primordial $B$-mode reconstruction and get prepared to the challenges that we are going to face with the next generation of CMB experiments. The current lack of data represents an important limitation to achieve this goal, which is not going to be solved in the near future, as the high sensitivity/high resolution foreground observations will be carried out at the same time as the CMB one, with the coming multi-frequency observatories. \par

In this work, we propose  to overcome the current limitation of foreground models taking advantage of Generative Adversarial Neural Networks (GANs), with the goal of simulating realistic non-Gaussian polarized Galactic emission at sub-degree angular scales. In particular, we show how GANs can be trained to understand the underlying relation between large (degree) and small (tens of arc-minutes) angular scales, when reliable data  are available. The trained network can be applied to reproduce this relation and the statistical properties of small scales whenever data are not available. Very recently other works have applied generative models to simulate thermal dust maps using either Variational Auto Encoders (VAE) \citep{2021arXiv210111181T} or through a statistical denoising method based on the wavelet phase harmonics statistics \citep{2021arXiv210203160R}.\par

The paper is organized as follows: in Section \ref{Sec:FG} we summarize the current knowledge of Galactic foreground emission in the context of CMB observations, and the status of available models with their limitations. In Section \ref{sec:forse} we introduce GANs and we describe \forse (Foreground Scale Extender): the algorithm developed in the context of this work. In Section \ref{sec:results} we show the results obtained when the proposed method is applied to thermal dust emission in both total intensity and polarization.  Our conclusions together with the description of the numerous possible future applications of our approach are summarized in Section \ref{sec:conclusions}.

\section{Galactic contamination to CMB observations}
\label{Sec:FG}
In this Section we summarize the current knowledge of Galactic emission as contaminant to CMB observations and we describe the available Galactic models that are currently being used in preparation of the future CMB experiments. 

\subsection{Status of observations}
The microwave sky is characterized by the presence of two main Galactic emissions that generate highly linearly polarized radiation. \par

At low frequencies ($\lesssim70$ GHz) synchrotron emission from cosmic ray electrons accelerating around the Galactic magnetic field represents the dominant source of radiation. Synchrotron polarized emission has been observed on the full sky by both the WMAP and Planck satellites, with the highest signal-to-noise ratio (SNR) reached at frequencies of 23 and 28.4 GHz respectively. The Stokes $Q$ and $U$ maps obtained from these observations have allowed to constrain the emission with sufficient sensitivity at angular scales larger than few degrees \citep{2016A&A...594A..25P}. Other datasets, with partial sky coverage but with higher SNR, have been obtained from observations at lower frequencies (S-PASS \citet{2019MNRAS.489.2330C, 2018A&A...618A.166K}, C-BASS \citet{2018MNRAS.480.3224J}).  Thanks to these observations we know that synchrotron emission is highly linearly polarized, with polarization fraction that can reach about 20\% at intermediate and high Galactic latitudes. In the first approximation the synchrotron SED (spectral energy distribution) follows a power law  $A_s(\nu)\propto\nu^{\beta_s}$ with spectral index $\beta_s\approx-3$ that presents a non negligible spatial variation \citep{2020MNRAS.495..578J, 2018A&A...618A.166K, Fuskeland14}. At the power spectrum level, synchrotron amplitude can be approximated with a power law as a function of multipole with  $A_s(\ell)\propto\ell^{\alpha_s}$ and $\alpha_s\approx-3$, showing therefore a pretty steep decay at small angular scales  \citep{2018A&A...618A.166K}. \par

At higher frequencies ($\gtrsim100$ GHz) the polarized thermal emission coming from dust grains aligned with the Galactic magnetic field represents the prevalent radiation in the microwave sky. Polarization maps of thermal dust emission have been obtained from Planck observations on the full sky at the frequency of 353 GHz with an angular resolution of about 5 arc-minutes. These maps are signal dominated everywhere in the sky at angular scales larger than $\sim1^{\circ}$, while finer structures are detected with high accuracy only at low Galactic latitudes. Dust radiation in Planck data is well fitted by a single modified black body SED with a spectral index $\beta_d\approx1.6$ and a temperature $T_d\approx20$ K \citep{PIP-XXX, 2020A&A...641A..11P}. Both these spectral parameters show variations in the sky on the degree scale \citep{planck2015-X, 2016A&A...596A.109P}. Planck observations have also shown that there exists an asymmetry in the polarized dust radiation, with $E$-mode power being about twice as higher than $B$-modes \citep{PIP-XXX, 2016A&A...586A.141P}. Similarly to synchrotron radiation, the thermal dust power spectrum can be approximated at first order with a power law  $A_d(\ell)\propto\ell^{\alpha_d}$ , with  ${\alpha_d}\approx-2.4$. The dust signal also show high level of non-gaussianity with non-zero polarized bispectra (the equivalent of the three point correlation function in harmonic space) that have been observed in the Planck maps \citep{2019JCAP...10..056C}. \par
Polarized synchrotron and thermal dust radiation have a high degree of spatial correlation, due to the same underlying Galactic magnetic field. In particular, the joint analysis of Planck and WMAP data allowed to measure the correlation coefficient $\rho$ which shows a progressive decay as a function of multipole, with $\rho\approx0.5$ at $\ell\approx10$ and approaching values compatible with zero at the degree angular scales \citep{2020A&A...641A..11P, Choi15}.

\subsection{Available foreground models}
\label{sec:fg_models}
Given the brightness of polarized Galactic emission and its complexity having reliable models of foregrounds is fundamental  to be able to optimize and test the capability of component separation algorithm to retrieve clean CMB maps. In fact, in the recent years a mis-modeling of thermal dust emission has already led to a false detection of primordial $B$-modes from the observations of the BICEP2 experiment \citep{BICEP2, PIP-XXX, B2P}. \par
Current models are largely based on foreground templates obtained from the datasets described in the previous section. These models are included in dedicated packages that allow to simulate the microwave sky in the different frequency channels (e.g. the Python Sky Model, PySM\footnote{The PySM package, its documentation and a description of the implemented sky models is available here \url{https://pysm3.readthedocs.io/en/latest/}}, \cite{2017MNRAS.469.2821T}; the Planck Sky Model, PSM, \cite{2013A&A...553A..96D}).  As it has been already stressed, these templates have a close match to the real sky at angular scales $\gtrsim1^{\circ}$ but cannot be used to simulate smaller structures, since at those scales observations are largely contaminated by instrumental noise. This limitation is usually overcome by extrapolating foreground power spectra at higher multipoles considering a simple power law models. Gaussian realizations of these spectra are then generated and combined with the large scale templates in order to obtain foreground maps with the inclusion of small scale structures.  Although this procedure can guarantee to have, at least at the first order, foreground simulations with the correct amplitude at all scales, it obviously represents quite a rough simplification from a statistical point of view. In fact, the higher order statistics of foregrounds must be taken into account to estimate its impact on both lensing reconstruction and on possible detection of primordial non-Gaussianity. New, more sophisticated Galactic models are therefore necessary in order to achieve this goal. \par
A second approach to model the Galactic emission is based on numerical magnetohydrodynamic (MHD) simulations. In this case realistic magnetized interstellar medium structure is simulated taking into account turbulence driven by star-formation and supernova feedbacks. Synthetic foreground maps can be generated through MHD simulations, and it has been shown that properties of real Galactic emission can be retrieved, as for example the asymmetry in the amplitude of polarized $E$-modes compared to the $B$ ones observed in thermal dust radiation (see for example \cite{2019ApJ...880..106K} and references therein). However with this approach is difficult to simulate maps with large scale structures corresponding to the real foreground morphology and for this reason MHD simulations are currently little used in CMB studies. \par

\citet{Vansyngel2018} proposed a hybrid approach, combining magnetic field informations with the latest Planck observations. They aimed at characterizing the small scales of thermal  dust polarization  by relating them  to the underlying  structure of the Galactic magnetic field, modeled as  a  uniform field plus  a random turbulent component. The parameters of the magnetic field model  are constrained so that they could reproduce the  dust EE, BB and TE power spectra observed  in  Planck maps. Their simulated maps  are produced at an angular resolution of $0.5^{\circ}$.

\section{The \forse algorithm}
\label{sec:forse}

\begin{figure*}
\centering
\includegraphics[width=14 cm]{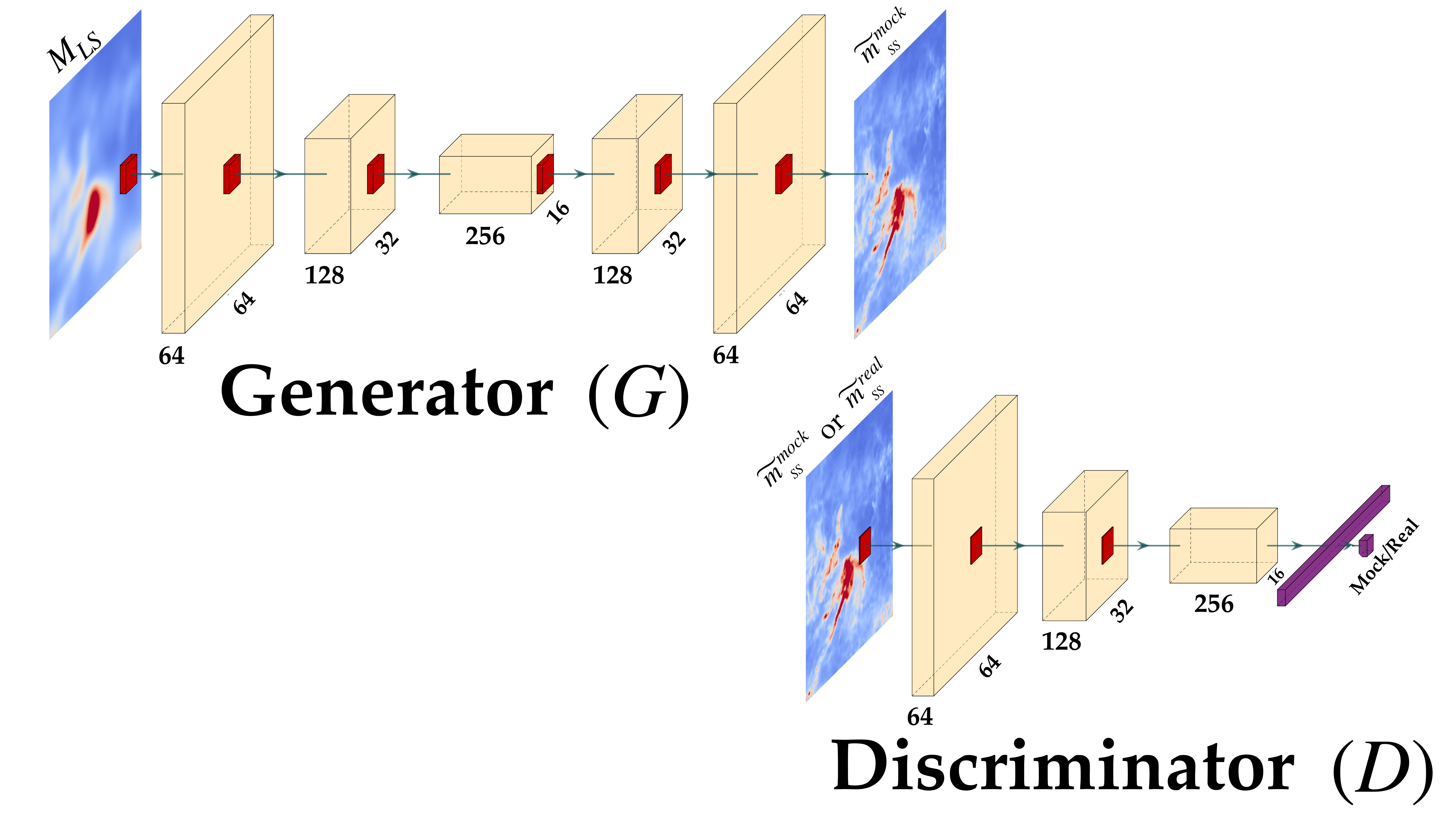}
\caption{\label{fig:arch} Sketch of the GAN architecture developed for this work. The \textit{Generator} ($G$) takes as input images of $320\times320$ pixels, which represent patches of the sky where only large scale structures are present  ($ M _{LS}$). The output of $G$ are mock images of the small scale features ($\widetilde{m }_{_{SS}}$). The \textit{Discriminator} ($D$) network takes as input real or mock small scale structures and classify them. The red squares represent the convolutional kernels (with dimension $5\times5$). A detailed description of the different layers is given in Section \ref{sec:arch}.}
\end{figure*}

In this work, we propose a new approach to generate simulated maps of the polarized Galactic emission at sub-degree angular scales. The package that we developed, called \forse (Foreground Scale Extender), is based on the use of Neural Networks (NNs) to generate realistic small scale structures in foreground maps\footnote{The word ``forse'' means ``maybe'' in Italian. With this name we emphasize the fact that the small scale structures generated by our algorithm are not expected to reproduce the true morphology of foregrounds but just to predict realistic features with the correct statistical properties. }. Our NN architecture, the weights of the trained networks, as well as all the codes needed to reproduce our results are publicly available\footnote{\url{https://github.com/ai4cmb/ForSE}} \citep{forse_code}.\par
NNs have started to be widely used in Astrophysics and Cosmology, with several applications also in the CMB field: from parameter estimation \citep{2019A&A...628A.129K} to lensing reconstruction \citep{2019A&C....2800307C}, foreground characterization and component separation \citep{2020arXiv200411507P, 2020JCAP...07..017F}.\par
In this section we summarize how GANs, the family of NNs used in this project, work and we describe the architecture used for our application.

\subsection{Generative Adversarial Networks (GANs)}
The GAN approach was firstly  proposed by \citet{Goodfellow2014}  and since then it has become more and more popular in several contexts where a generative framework can be applied, e.g. in natural image processing as image inpainting, super-resolution, image de-noising. To date, there are several GANs proposed  in the literature and the major differences can mostly be found in the network architecture  implementations (e.g. see Deep-Convolutional GAN (DCGAN)  \citealt{DBLP:journals/corr/RadfordMC15}, Super-Resolution GAN \citealt{srgan}, Contextual-Attention \citealt{Yu2018})\par 

The general idea of GAN framework relies on an \emph{adversarial game} between two NNs: a \emph{Generator}, $G$, and a \emph{Discriminator}, $D$. Practically, $G$ generates new data instances that must match the general feature of a given data sets (the training set); $D$ evaluates the authenticity of the samples generated by $G$,  distinguishing whether each sample belongs or not to the actual training set. $D$ returns a probability value between 0 and 1, with 1 (0)  labeling an authentic (mock) image. The goal of $D$ is thus to label all the generated images as ``mock'', whereas the goal of $G$ is to improve the quality of the generated images in order to fool $D$.\par

The whole process of training a GAN therefore consists of iterating between two feedback loops: $G$ being in a feedback loop with $D$, and $D$ in loop with the authentic set of images. Further details about our GAN implementation can be found in Section \ref{sec:ourgan}.\par

During the past few years, GAN have been adopted on a wide range of applications in cosmology.  \citet{2020MNRAS.495.4227K} and \citet{Mustafa17} used GANs in order to generate high-quality samples of respectively  N-body simulations and weak lensing convergence maps. \citet{Aylor2019}  generated  thermal dust intensity   emission maps with  an effective non-Gaussian statistical model     and  trained  a DCGAN  architecture  from  the observations  inferred  by the  \emph{Planck} satellite. \citet{puglisi2020} showed that  GANs can be efficiently applied to Galactic foreground unpolarized and polarized emission in order to generate emission at small scales and  in-paint the maps on regions smaller than $\lesssim 0.5 $ degrees.

\subsection{The implemented GAN}
\label{sec:ourgan}

The problem that we want to address can be summarized as follows. Let $M$ be a map encoding foreground emission at a given angular resolution. $M$ can be expressed as the sum of two maps encoding large and small scale structures, respectively $ M _{LS}$ and $M_{SS}$:

\begin{equation}
M = M_{LS} + M_{SS}.
\end{equation}
We can also make the assumption that the small scale features are actually modulated by the large ones, with $M_{SS}=M_{LS}\cdot m_{_{SS}}$. In this way we have: 

\begin{equation}
\label{eq:m_ss}
\widetilde{m }_{_{SS}}\equiv m_{_{SS}} + 1 = \frac{M}{ M_{LS}}.
\label{eq:Mm}
\end{equation}

The goal of this work is to use GANs to generate a map of realistic small scale structures ($\widetilde{m }_{_{SS}}$) given the real large scale ones.

\subsubsection{Architecture}
\label{sec:arch}

The \forse package is based on a modified version of the DCGAN described in \cite{DBLP:journals/corr/RadfordMC15}. In its original version the \textit{Generator} ($G$) of a DCGAN takes as input a vector of random numbers and, by reshaping it and applying a series of convolutional layers, generates a mock image. However, our application is different, as both the input and the output of $G$ are images.\par
The GAN architecture implemented in this work is sketched in Figure \ref{fig:arch}. $G$ takes as input images of $320\times320$ pixels which are maps (patches) of the foreground sky where only the large scale features are present ($M_{LS}$). A series of three convolutional layers is then applied with the dimension of the kernel being $5\times5$. In the first layer $64$ filters are used, and no stride is applied, leading to an image cube with dimension of $320\times320\times64$ pixels. In the following convolutional layers a stride of $2$ is used and the number of filters is doubled, generating therefore cubes of  $160\times160\times128$ and $80\times80\times256$ pixels after the second and third convolution, respectively. After each convolution, a \textit{LeakyRelu} activation function is applied with slope equal to $0.2$, and a \textit{BatchNormalization} layer is added. \par
The decoding part of $G$ is symmetric to the encoding one; upsampling layers are combined with convolutional ones in order to restore the output dimension of the image with $320\times320$ pixels (see Figure \ref{fig:arch}). The output layer is activated with a \textit{tanh} function. \par
The \textit{Discriminator} ($D$), also shown in Figure \ref{fig:arch}, takes as input  $\widetilde{m }_{_{SS}}$ images of $320\times320$ pixels. After three convolutional layers (which are analogous to the encoding part of $G$) the resulting cube is flattened into a 1-D vector which is then densely connected to the output unit, activated though a \textit{sigmoid} function. \par
The GAN architecture and the training procedure (described in the following section) has been developed using the \textit{Keras}\footnote{See also \url{https://keras.io/api/}  for a thorough description of the layers and functions used in our GAN.} python library with a \textit{tensorflow} backend. 

\subsubsection{Training procedure}
\label{sec:train}
In all our applications we trained our GAN by minimizing a \textit{binary cross-entropy} loss function applying a stochastic gradient descent. As suggested in   \cite{DBLP:journals/corr/RadfordMC15}, we used the \textit{Adam} optimizer, with a learning rate of $0.0002$ and a momentum term $\beta_1=0.5$.  \par
The training was done considering mini-batches of $N_{b} = 16$ images with the following two steps:
\begin{enumerate}
\item   $N_{b}$ patches of the real large scale structure ($M_{LS}$) were fed to $G$ which generates the corresponding $\widetilde{m }_{_{SS}}^{mock}$ and passes them to $D$. In this phase $G$ was trained by maximizing the probability that $\widetilde{m }_{_{SS}}^{mock}$ are classified as real images (class $1$) by $D$. The back-propagation is done trough the whole GAN ($G$+$D$) but only the weights of $G$ are optimized.

\item In the second step $N_{b}/2$ $\,\,\widetilde{m }_{_{SS}}^{mock}$ and $N_{b}/2$ $\,\,\widetilde{m }_{_{SS}}^{real}$ images were passed to $D$ which was trained to classify them as mock (class $0$) and real ($1$), respectively. 
\end{enumerate}  
As a pre-step, all the $M_{LS}$ and the $\widetilde{m }_{_{SS}}^{real}$ patches were normalized, in order to have pixel values in the range $[-1, 1]$.

\section{Application to thermal dust emission}
\label{sec:results}
As described in Section \ref{Sec:FG}, thermal dust emission represents one of the strongest contaminant to CMB observations  at frequency around 150 GHz, and shows high level of non-Gaussianity in both total intensity and polarization.  We therefore choose to apply our algorithm to thermal dust maps, with the goal of obtaining simulations with realistic and non-trivial small scale structures at sub-degree angular resolution. In this Section, we describe the methodology used as well as the results obtained.  \par

Existing templates of thermal dust emission on the full sky have been obtained by applying component separation algorithms to Planck maps. In this work, we used the ones obtained from the GNILC (Generalized Needlet Internal Linear Combination) method at 353 GHz, which have the advantage of not being contaminated by the CIB radiation in total intensity \citep[for a detailed description of the data and the method applied to get this thermal dust template we refer to][]{2016A&A...596A.109P}. The GNILC dust maps have an angular resolution that varies in the sky and depends on the SNR of the Planck high frequency maps: in total intensity the effective beam FWHM (full width half maximum) ranges from $5$ to about $22$ arc-minutes, while in polarization it varies in the interval $5-80$ arc-minutes \citep{2020A&A...641A...4P}\footnote{The GNILC maps are available on the Planck Legacy Archive website \url{http://pla.esac.esa.int/pla/\#maps}}. Both polarization and intensity maps follow the HEALPix\footnote{\url{https://healpix.sourceforge.io/index.php}} pixelization scheme \citep{2005ApJ...622..759G} at $N_{side}=2048$, i.e. $\sim 1.7 $ arc-minute  pixel resolution. 
 
\subsection{Total intensity}
\label{sec:intensity}

\begin{figure}[!tb]
\centering
\includegraphics[width=8 cm]{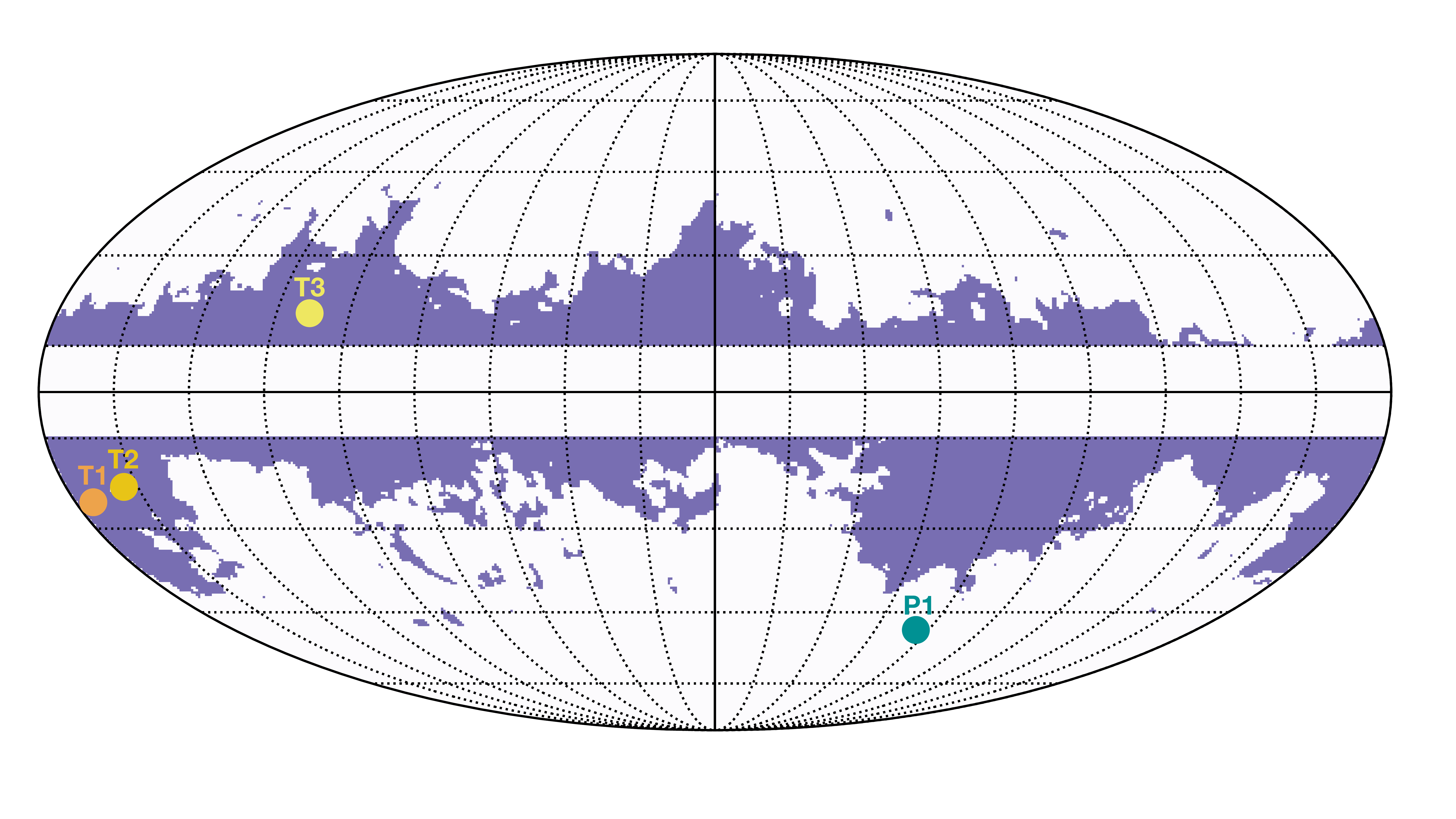}
\caption{Sky mask adopted in this work (see {Sect. \ref{sec:intensity}  } for details). Images used to perform the GAN training have been selected within the purple region. Yellow dots represent the position of the center of the patches shown in Figs. \ref{fig:intensity_inout}-\ref{fig:intensity_spectra}, whereas the green one corresponds to the patch in Figs. \ref{fig:QU_inout}-\ref{fig:pol_patches}. \label{fig:mask}}
\end{figure}

We firstly tested the ability of \forse of generating realistic small scale features in total intensity (quantified by the Stokes $I$  parameter). To do so, we needed a set of patches for which the real small scale structures have been observed. As stated above, the GNILC template in total intensity has a variable angular resolution, which is equal to 5 arc-minutes in the regions close to the Galactic plane thus making them suitable to be used to train the GAN. In order to further select only those regions less contaminated by noise, we built a mask from the Planck HFI 353 GHz $I$ map. In particular, we took into account only that part of the sky where the SNR in the 353 GHz map at the full angular resolution (about 5 arc-minutes) is above a threshold of 8. We smoothed the obtained mask with a Gaussian beam with FWHM of $1^{\circ}$, in order to regularize its borders. We also cut out the inner part of the Galactic plane (with $b<10^{\circ}$) which, due to its peculiarity, is not suitable to be used as part of the training set. The resulting mask is shown in Figure \ref{fig:mask} and covers a fraction of the sky of about  23\%. 

In order to train the GAN, both  $M_{LS}$ and $\widetilde{m }_{_{SS}}^{real}$ images were needed.  The  $M_{LS}$ patches were generated as tiles of $20^{\circ}\times20^{\circ}$ and $320\times320$ pixels from the GNILC $I$ template, smoothed at the angular resolution of 80 arc-minutes. On the other hand, $\widetilde{m }_{_{SS}}^{real}$ patches were obtained from the same map smoothed at 12 arc-minutes angular resolution and divided by $M_{LS}$ (see definition in Eq. \ref{eq:m_ss}). All the tiles were taken from the region of the sky within the mask in Figure \ref{fig:mask},  for a total of 350 pair of images.
We trained our GAN on NVIDIA V100 GPUs, with the procedure outlined in Section \ref{sec:train}, for about $10^5$ iterations, saving both the $G$ and $D$ weights every 1000 steps. As described below, we obtained the best performance after 37000 iterations, and therefore, in the following, we report the results corresponding to this GAN configuration.  \par

\begin{figure*}[!htb]
\centering
\includegraphics[width=12 cm]{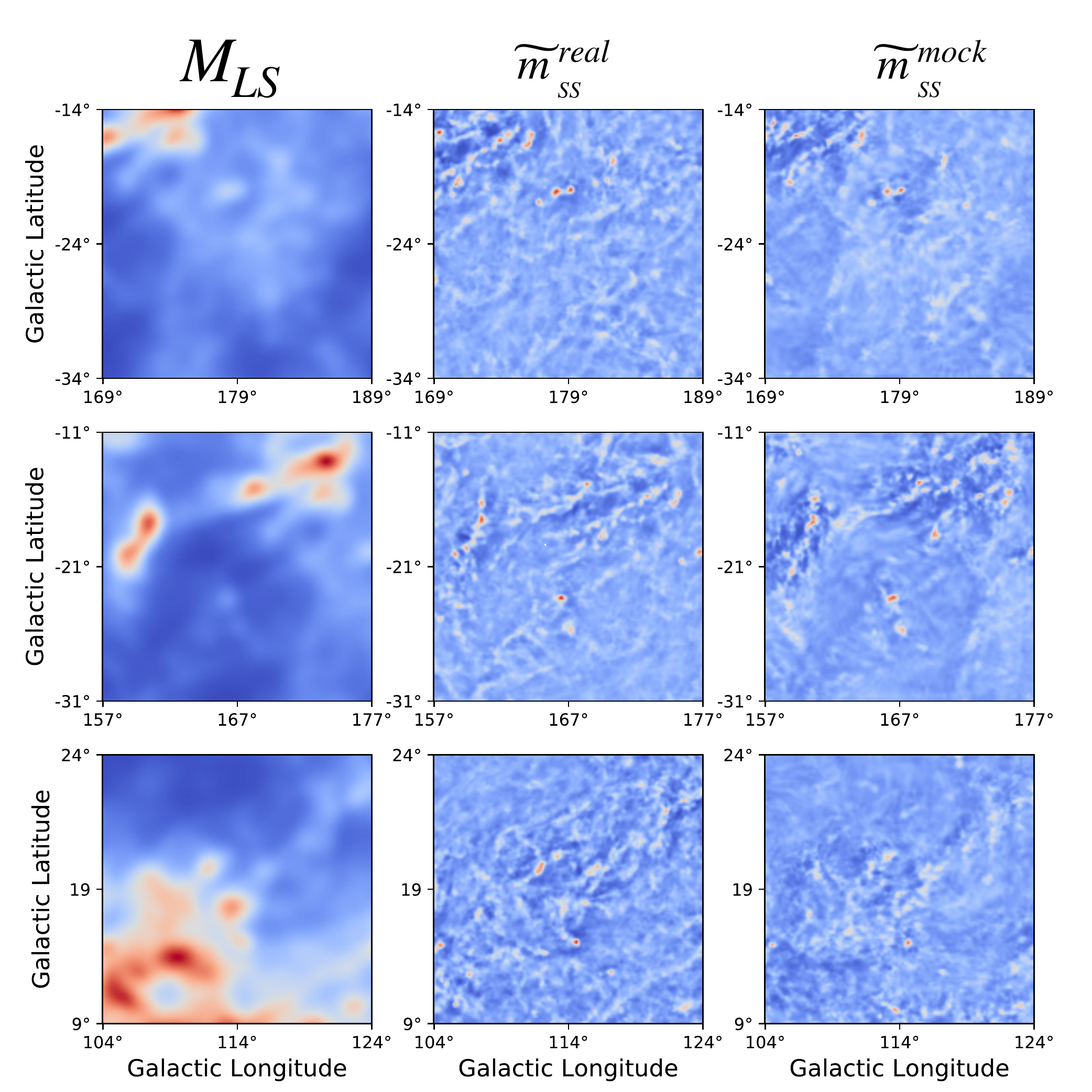}
\caption{\label{fig:intensity_inout} Input (left column) and output (right column) of \forse when applied to total intensity dust map for three different sky regions. The middle column shows the real small scale features (12 arc-minutes) present in the sky. Maps are rescaled to unphysical units ranging  in the same interval $[-1, 1]$. The position of the three patches in the sky is reported with yellow dots in Fig. \ref{fig:mask} (T1 to T3 from upper to lower panel). Note that patch T1 and T2 partially overlap, showing how the GAN generates similar small scale structures in the overlapping region.}
\end{figure*}

\begin{figure*}[!htb]
\centering
\includegraphics[width=16.4 cm]{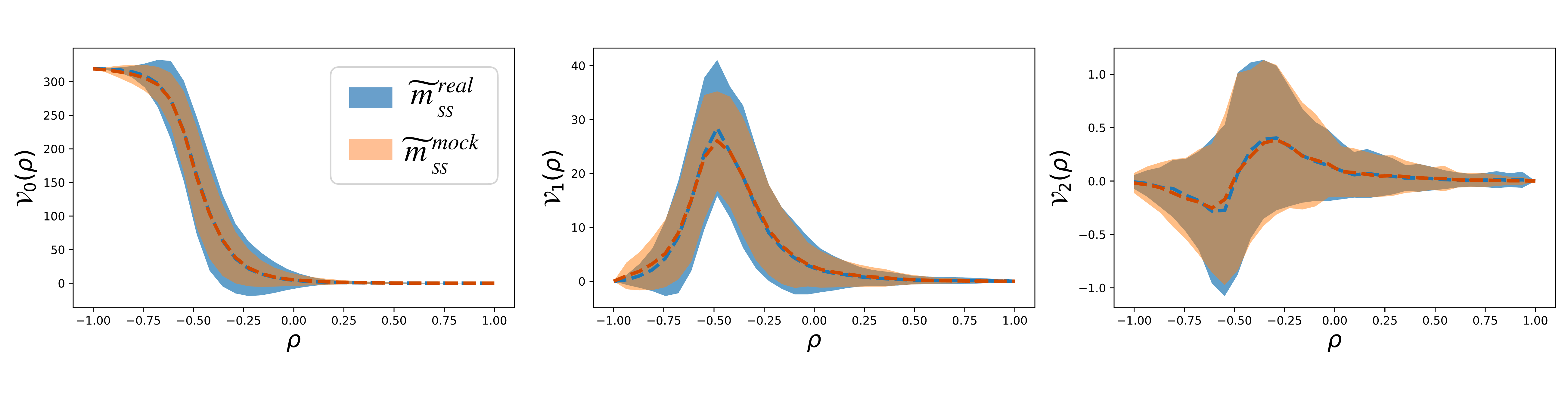}
\caption{\label{fig:intensity_minko} Minkowski functionals ($\mathcal{V}_0$, $\mathcal{V}_1$, $\mathcal{V}_2$) as a function of the threshold $\rho$, for $\widetilde{m }_{_{SS}}^{real}$ (blue) $\widetilde{m }_{_{SS}}^{mock}$  (orange) for the case of total intensity dust maps. The functionals are computed for the 350 patches used to train the GAN and we report the mean (dashed lines) and $1\sigma$ deviation (shaded areas) of the distributions. The high level of superposition between the curves demonstrates the ability of \forse to reproduce the correct statistical properties of small scale foregrounds.}
\end{figure*}

Figure \ref{fig:intensity_inout} reports the input and output of $G$ after training, as well as the comparison with the real small scale features, for three different sky location. Images are normalized in the range $[-1, 1]$. These results show the capability of  \forse of generating small scale features with a non-trivial statistics, starting only from the large scale ones. We stress again that, with this approach, we do not expect to generate small scale structures that reproduce the morphology of the real ones, but to retrieve the correct statistics. It is however remarkable how the network is able to reconstruct the main small scale features (e.g. peaks and troughs), implying that during training it has learned the underlaying relation between large and small scales. \par

In Figure \ref{fig:intensity_minko} we show a comparison of the generated small scale structures with the real ones, from a statistical point of view. In order to do so we made use of the Minkowski functionals $\mathcal{V}_0$, $\mathcal{V}_1$, $\mathcal{V}_2$, as defined in \cite{Mantz08}, which are sensitive to the presence  of non-Gaussian structures. The three functionals characterize the area, the perimeter and the connectivity of the features respectively. We computed $\mathcal{V}_0$, $\mathcal{V}_1$, $\mathcal{V}_2$ as a function of the threshold $\rho$ in the range $[-1, 1]$, for all the 350 patches and for $\widetilde{m }_{_{SS}}^{real}$ and  $\widetilde{m }_{_{SS}}^{mock}$; in Figure \ref{fig:intensity_minko}, for each functional, we plotted the mean and $1\sigma$ deviation. The two distributions present a remarkable agreement, with superposition at the level of 76\% ($\mathcal{V}_0$), 84\% ($\mathcal{V}_1$) and 91\% ($\mathcal{V}_2$). We have computed these values for all the different GAN configurations saved during training, and chose the one that led to the best agreement (reached after 37000 iterations). This results clearly show how the approach developed with \forse allows to generate small scale feature on foreground maps that match the statistical properties of the real ones. \par
Once the small scale features are generated by the GAN they needed to be normalized back in order to restore physical units. In this case, where we tested the feasibility of the approach in total intensity, the normalization to physical units is trivial, as, for each considered patch, we know the amplitude of the real small scales structures and  we could therefore use this information. In practice, we rescaled each $\widetilde{m }_{_{SS}}^{mock}$ patch in order to match its mean and standard deviation with the ones of the corresponding $\widetilde{m }_{_{SS}}^{real}$. Once physical units were restored, we combined the large and small scale patches and got the final image as $M = M_{LS} \cdot \widetilde{m }_{_{SS}}$ (see relation \ref{eq:Mm}). \par

\begin{figure*}[!htb]
\centering
\includegraphics[width=18 cm]{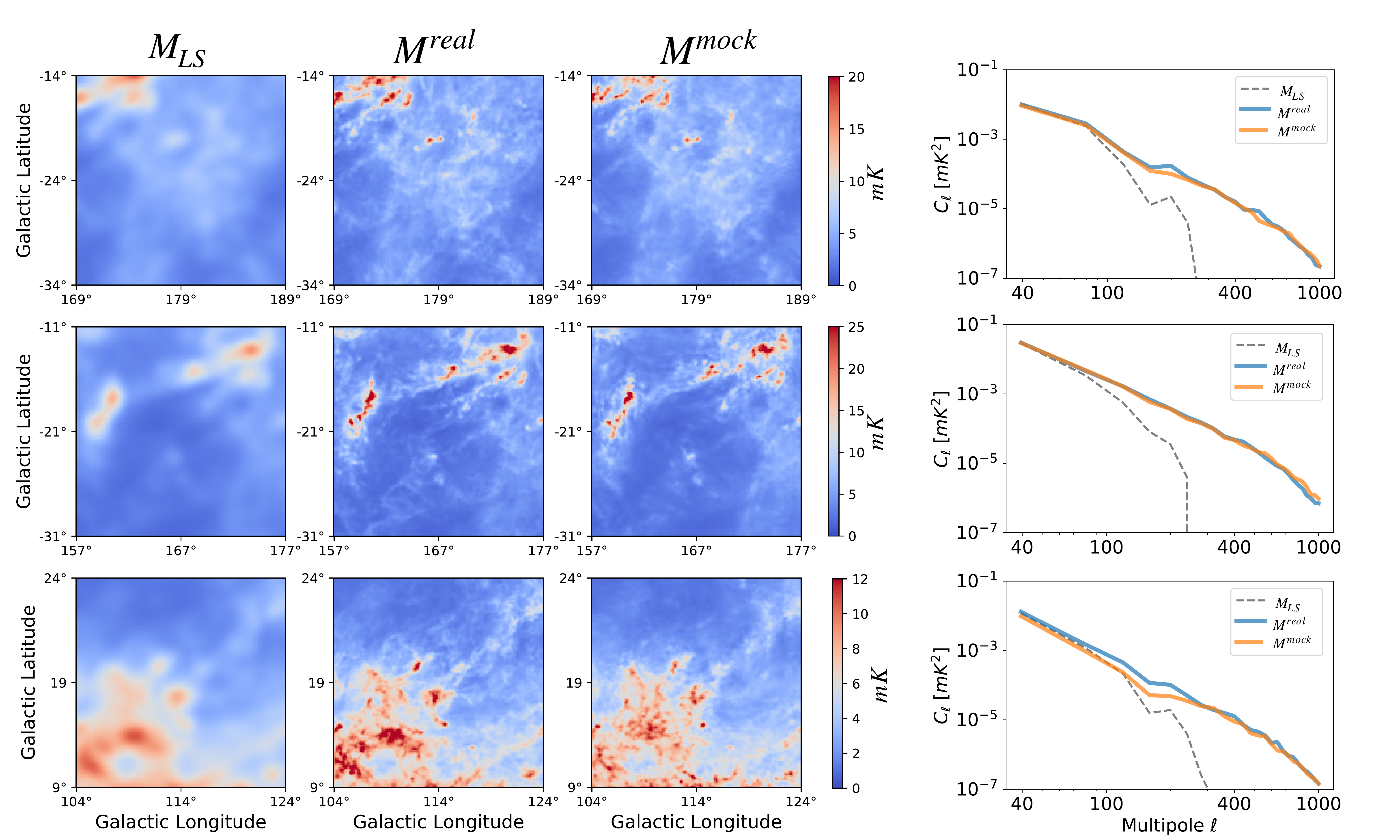}
\caption{\label{fig:intensity_spectra} Left panel: large scale ($M_{LS}$) and combined ($M^{real/mock}$) images for the three different patches shown in Fig. \ref{fig:intensity_inout} as obtained after restoration of physical units. The position of the patches in the sky is shown with yellow dots on Fig. \ref{fig:mask}}. Right panel: angular power spectra of the images shown on the left. 
\end{figure*}

Figure \ref{fig:intensity_spectra} shows our final results. In the left panel, we report the large scale and combined images (large and small scales) for the same three patches shown in Figure \ref{fig:intensity_inout}, as an example. We report also their angular power spectra (right panel) computed with the NaMaster code \citep{2019MNRAS.484.4127A}. The comparison of the combined $M^{mock}$ and  $M^{real}$ images and spectra not only shows that \forse generates small scale structures with realistic morphology and statistical properties, but also that the correct amplitude scaling as a function of the angular scale is recovered after the normalization to physical units.

\subsection{Polarization}
\label{sec:polarization}

\begin{figure*}[!htb]
\centering
\includegraphics[width=12.5 cm]{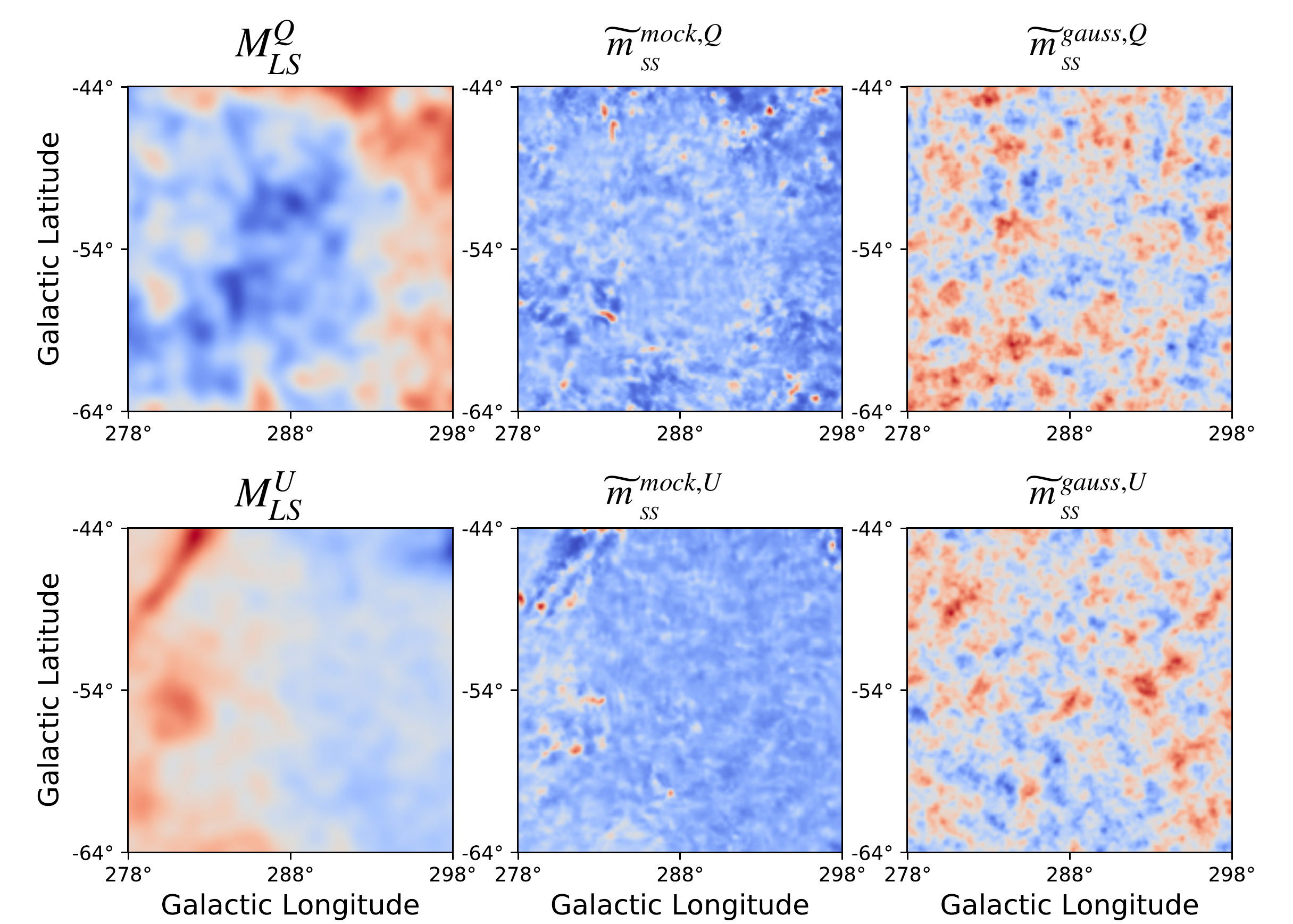}
\caption{\label{fig:QU_inout} Input (left column) and output (middle column) of \forse when applied to Stokes $Q$ and $U$ maps (upper and lower rows respectively). The small scale structures generated by the GAN are compared with Gaussian ones obtained as described in Section \ref{sec:polarization}. Maps are rescaled to unphysical units ranging in the interval $[-1, 1]$. The position of this patch in the sky is shown with a green dot (P1) in Fig. \ref{fig:mask}.}
\end{figure*}

\begin{figure*}
\centering
\includegraphics[width=15.5 cm]{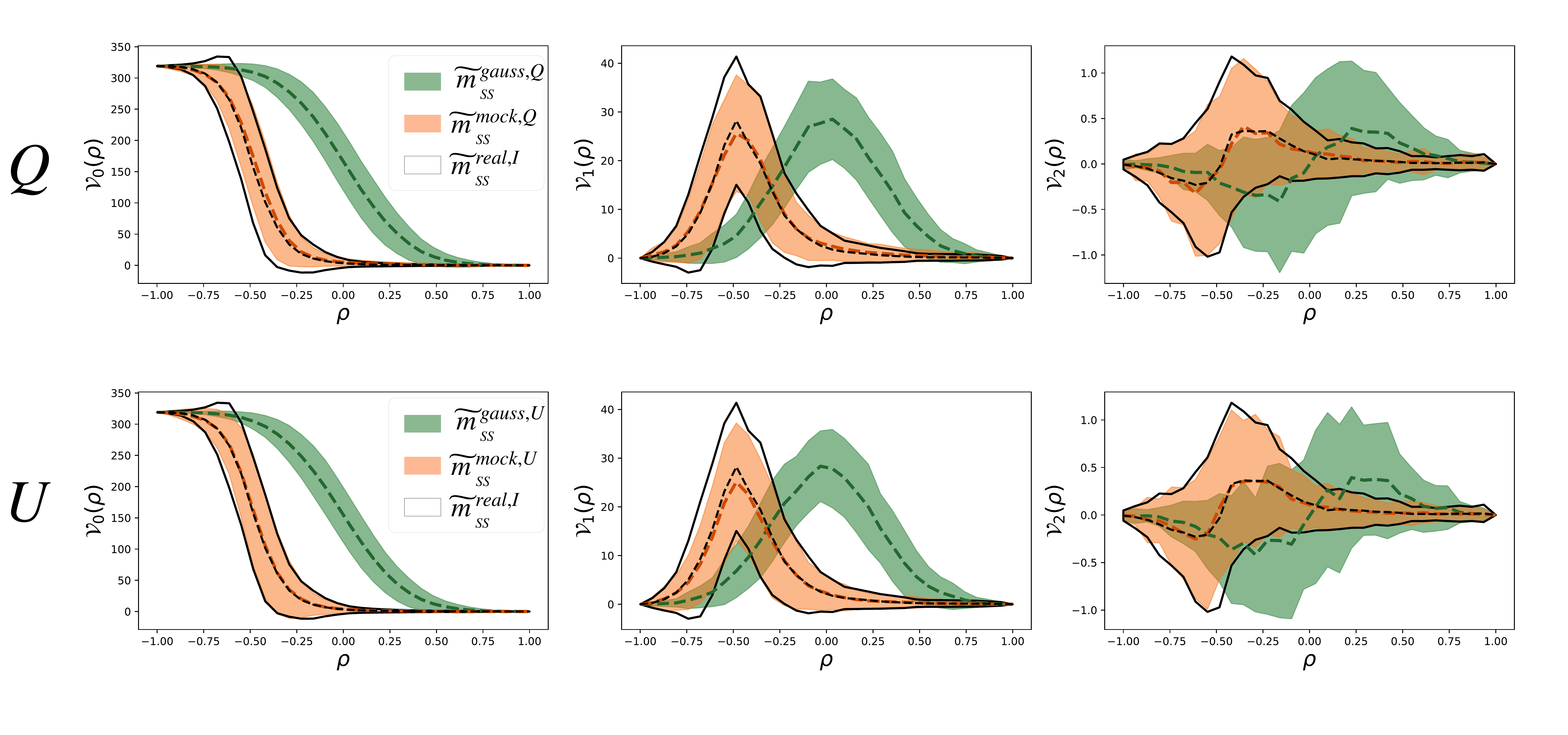}
\caption{\label{fig:pol_minko} Minkowski functionals for real small scales structures in total intensity (black lines), features generated by the GAN (orange) and Gaussian ones (green). The functionals are computed for the 174 patches used to train the GAN and we report the mean (dashed lines) and $1\sigma$ deviation (shaded areas) of the distributions.}
\end{figure*}

\begin{figure*}[!htb]
\centering
\includegraphics[width=17 cm]{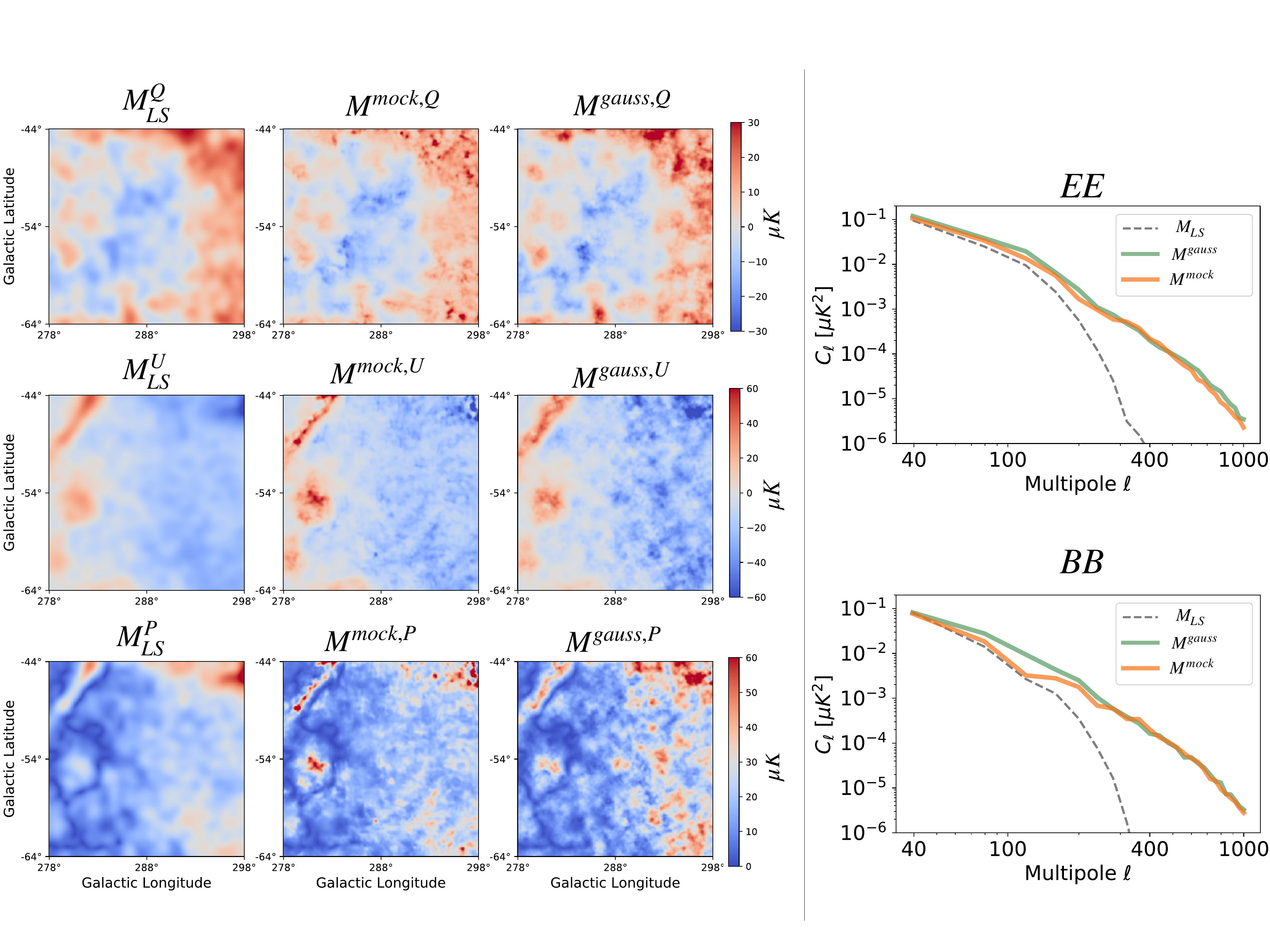}
\caption{\label{fig:pol_patches} Left panel: large scales ($M_{LS}$, right column) and combined images with small scales features obtained from the GAN ($M^{mock}$, middle column) or as Gaussian realizations ($M^{gauss}$, right column), for $Q$ (upper row), $U$ (middle row) and P (lower row). The sky region is the same as the one shown in Fig. \ref{fig:QU_inout} and corresponds to the green dot (P1) on the map in Fig. \ref{fig:mask}. Right panel: comparison of the $E$ (upper plot) and $B$-mode (lower plot) power spectra.}
\end{figure*}

After having tested our approach in total intensity, we could apply it to polarized emission. Although being more challenging, this represents the most interesting application of \forsee, as current CMB experiments are focusing on the observations of the polarized cosmological signal (see Section \ref{sec:intro}) and we suffer from a significant lack of foreground data.\par

The GNILC polarized thermal dust template has an angular resolution below 12 arc-minutes (the target resolution of \forsee) only in about 9\% of the sky and mainly in the inner Galactic plane region ($|b|<10^{\circ}$). Given this, we could not apply the procedure used in total intensity for Stokes $Q$ and $U$ maps, as we did not have enough data to perform training. To overcome this limitation, we made the assumption that small scale structures in polarization follow the same statistics as the ones in total intensity. This represents a reasonable assumption to the first order; in fact, we can presume that the dust grains  population producing the polarized emission  is the  same as the one responsible in emitting the unpolarized signal. Additionally, we already know that  thermal dust radiation has similar two point correlation functions (power spectra) in polarization and total intensity \citep{2020A&A...641A..11P}. \par

With this assumption, we proceeded by training separately two different GANs (both with the architecture described in Section \ref{sec:ourgan}), for Stokes $Q$ and $U$ patches respectively. In particular, we used as input to the GANs, images of the large scale structures $M_{LS}^{Q(U)}$ taken from the GNILC Stokes $Q$ ($U$) map, at the angular resolution of 80 arc-minutes. The \emph{Generator} $G$ was trained to generate small scale structures ($\widetilde{m }_{_{SS}}^{mock, Q(U)}$) at the angular resolution of 12 arc-minutes, starting from the large scale ones. These were then compared by the \emph{Discriminator} $D$ with the real small scale structures in total intensity ($\widetilde{m }_{_{SS}}^{real, I}$) and the GAN weights were thus optimized in order to make $\widetilde{m }_{_{SS}}^{mock, Q(U)}$ indistinguishable from $\widetilde{m }_{_{SS}}^{real, I}$. We applied the same training procedure used for total intensity (see Section \ref{sec:train}). \par

Since our goal was to obtain $Q$ and $U$ full sky maps, we generated the $M_{LS}^{Q(U)}$ tiles in such a way that they cover the full celestial sphere, accounting for a total of $174$ images for each of the two Stokes parameter  (the description of the procedure adopted to make the projections from the celestial sphere to the flat patches and back is reported in Appendix \ref{sec:appendix}). In order to get the best possible result, we optimized the GAN's weights directly on these images (after having normalized each of them in the range $[-1, 1]$) and compared the $G$ output with the $\widetilde{m }_{_{SS}}^{real, I}$ from the set of images used in the total intensity case. As before, and for both $Q$ and $U$ optimization, we trained the GAN for about $10^5$ iterations, saving  weights every 1000 steps. We reached the best performance (in terms of Minkowski functionals, see Figure \ref{fig:pol_minko}) after 83000 and 88000 iterations, for Stokes $Q$ and  $U$ maps respectively. \par

Figure \ref{fig:QU_inout} shows the results after training, for one patch and for both Stokes $Q$ and $U$ maps. The small scale structures generated by the GAN are compared with the ones from the Gaussian extrapolation of the dust power maps, obtained as follows: 
\begin{enumerate}
\item polarization ($E$ and $B$-modes) power spectra were computed from the GNILC $Q$ and $U$ full sky maps at the angular resolution of 80 arc-minutes and fitted with a power law in the range of multipoles $20-100$;
\item the fitted spectra were multiplied by a function $f_{\ell} = 1-B_{\ell}^2$ (where $B_{\ell}$ represents the beam window function with FWHM $= 80'$) in order to remove the large angular scales. A pair of $Q$ and $U$ maps was obtained as a Gaussian realization of them;
\item the obtained maps were multiplied by the GNILC $Q$ and $U$ maps at 80 arc-minutes in order to modulate the small scale structures, and then multiplied by a factor that restore the correct amplitude of the power spectra;
\item lastly the obtained small scale structure maps are summed to the GNILC large scale ones.
\end{enumerate}
This procedure generates maps which satisfy relation \ref{eq:Mm}, with Gaussian small scale structures $\widetilde{m }_{_{SS}}^{gauss}$, and whose power spectra follow the power law extrapolation of the real large scale features. We stress that a similar procedure is also implemented  in the publicly available foreground simulation packages (see Section \ref{sec:fg_models}).\par

Figure \ref{fig:pol_minko} shows the Minkowski functionals of the GAN-generated small scale structures for $Q$ and $U$ maps. The functionals are compared with the ones of the real structures in total intensity (which represent our target distribution), as well as with the ones computed from the Gaussian small scale features. The distributions of the mock structures are in good agreement with the real total intensity ones, with a superposition of the three functionals $(\mathcal{V}_0,\,\mathcal{V}_1,\, \mathcal{V}_2)$ at the level of $(79\%, 85\%, 89\%)$ and $(87\%, 84\%, 87\%)$ for $Q$ and $U$ maps respectively. On the contrary the distributions are very different from the Gaussian ones, showing once again how \forse is able to generate non-trivial structures on maps. \par
As for the total intensity case, we needed to normalize back the $\widetilde{m }_{_{SS}}^{mock, Q(U)}$ images to physical units. Since in polarization we do not have the information on the amplitude of the real small scale structure, we performed the rescaling by means of Gaussian extrapolations described above. Although these simulations present different statistical properties, they predict, at  a first order, the correct amplitude of the small scale structures (i.e. the extrapolation is specifically built in such a way that the spectra have the correct scaling). In practice, the $\widetilde{m }_{_{SS}}^{mock, Q(U)}$ maps were rescaled  in order to have the same mean and standard deviation as the corresponding Gaussian small scales structures in physical units. A second multiplication factor was applied in order to minimize the distance between the power spectra of the $\widetilde{m }_{_{SS}}^{mock, Q(U)}$ and $\widetilde{m }_{_{SS}}^{gauss, Q(U)}$ at $\ell>160$.\par

Figure \ref{fig:pol_patches} shows the final results, with the combination of real large scales and mock small ones for Stokes $Q$ and $U$ and polarized intensity $P=\sqrt{Q^2+U^2}$, together with the $E$ and $B$-modes power spectra. As it can be seen the GAN is able to produce non-Gaussian structures that after the normalization in physical units have the correct amplitude and scaling. However we notice that, given the fact that we trained the network to reproduce the statistics of total intensity features, it was unable to retrieve the thermal dust asymmetry of $E$ and $B$ modes. This limitation might be overcome by adopting two  different training sets  obtained   from   polarized intensity and polarization angle  patches, we outlook to explore this in the future, once high resolution polarization data will become publicly available. 

\subsubsection{Full sky maps}
\label{sec:fullsky}

\begin{figure*}
\centering
\includegraphics[width = 15.2 cm]{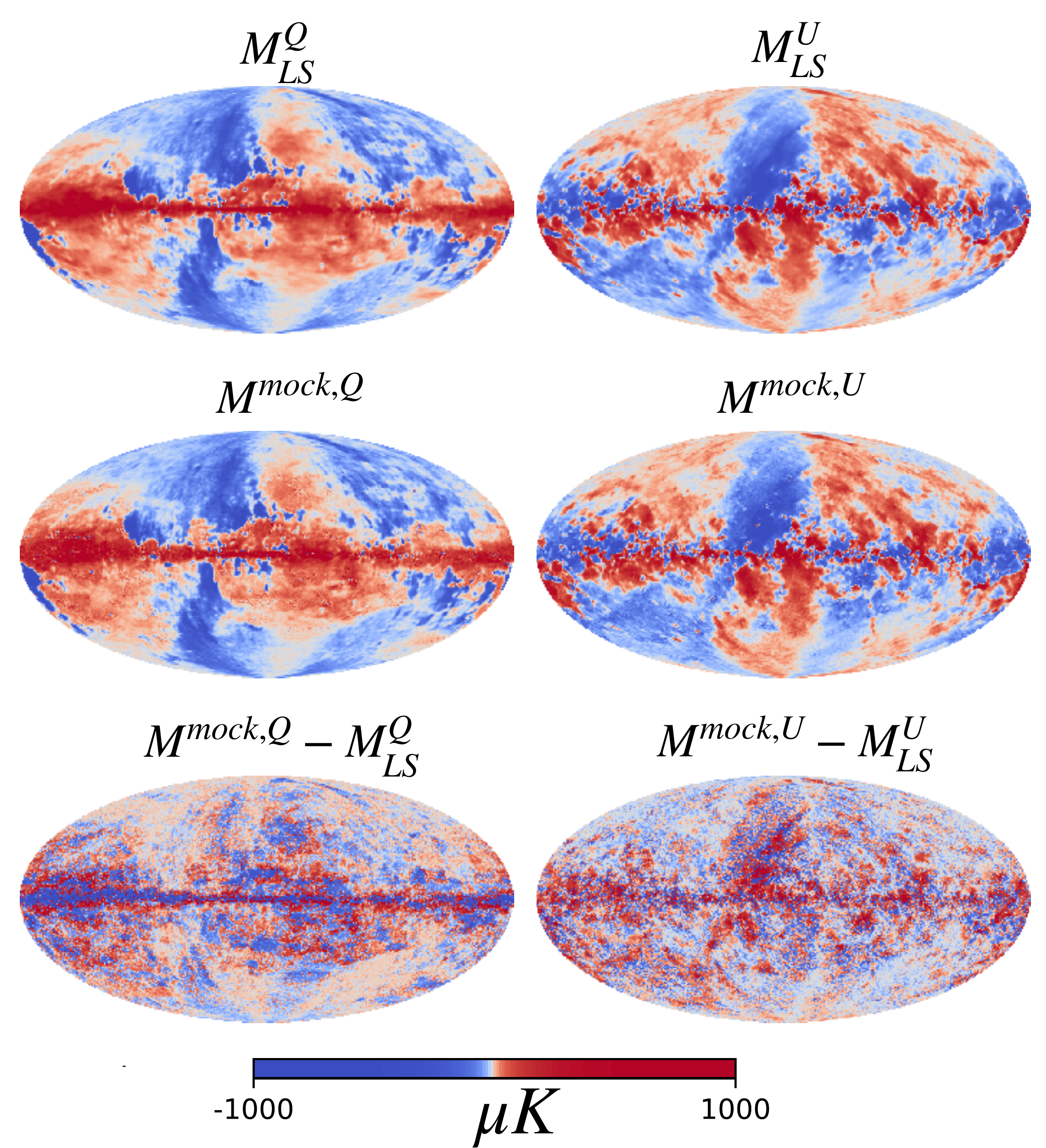}
\caption{\label{fig:QU_fullsky} Top row: full sky polarization maps (left: Stokes $Q$, right: Stokes $U$) for the GNILC template at 80 arc-mininutes angular resolution. These maps are  the input to our algorithm. Middle row: maps with small scale features, up to 12 arc-minutes,  generated by \forse. Bottom row: difference  between the two maps. Notice the residuals mostly encode smaller angular scales as expected.} 
\end{figure*}

\begin{figure}
\centering
\includegraphics[width = 8.7 cm]{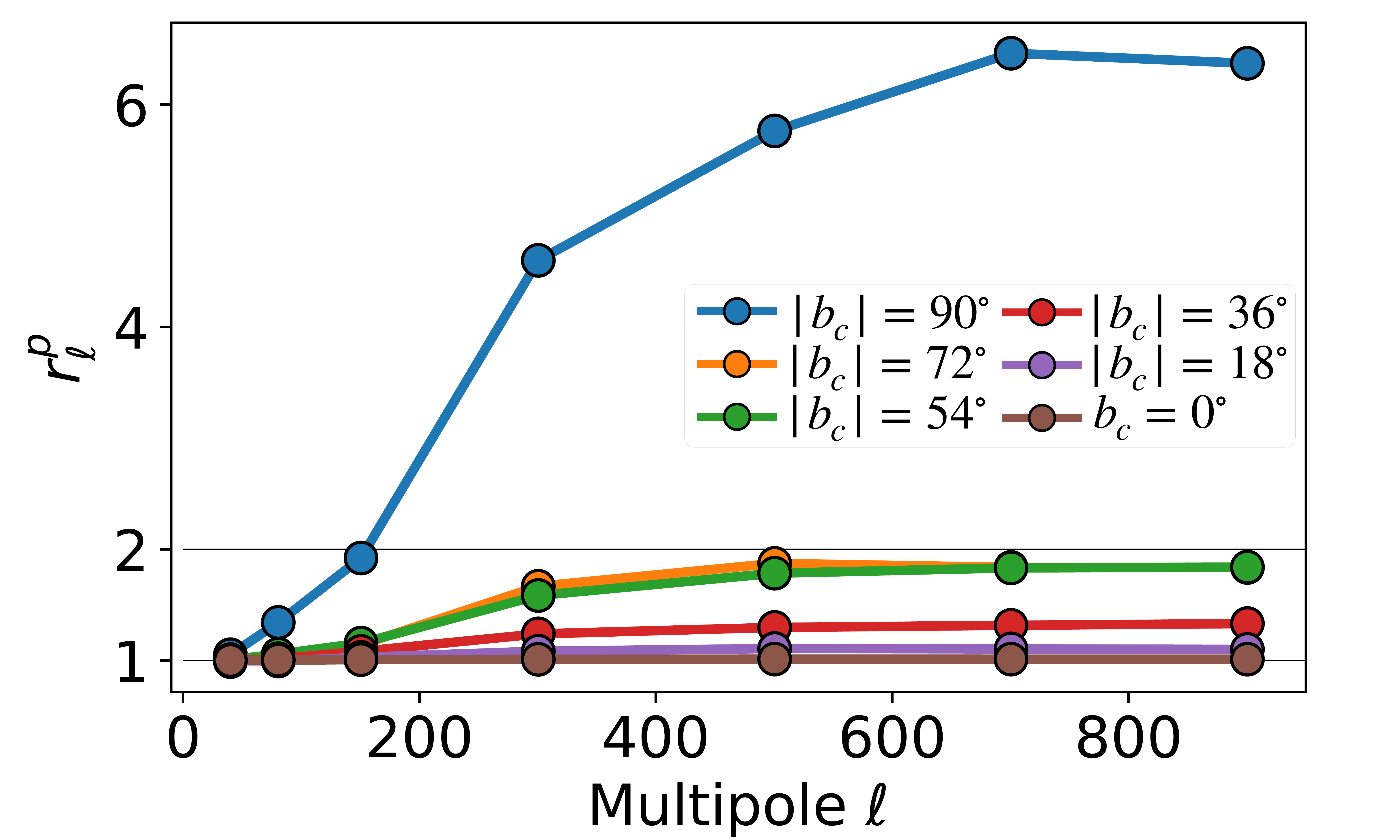}
\caption{\label{fig:reprj_eff}  Ratio between E-mode power spectra before and after reprojection averaged over patches located at the same absolute Galactic latitude. Smoothing effects due to the full sky reprojection appear as $r^p_{\ell}>1$. } 
\end{figure}

\begin{figure}
\centering
\includegraphics[width = 8.7 cm]{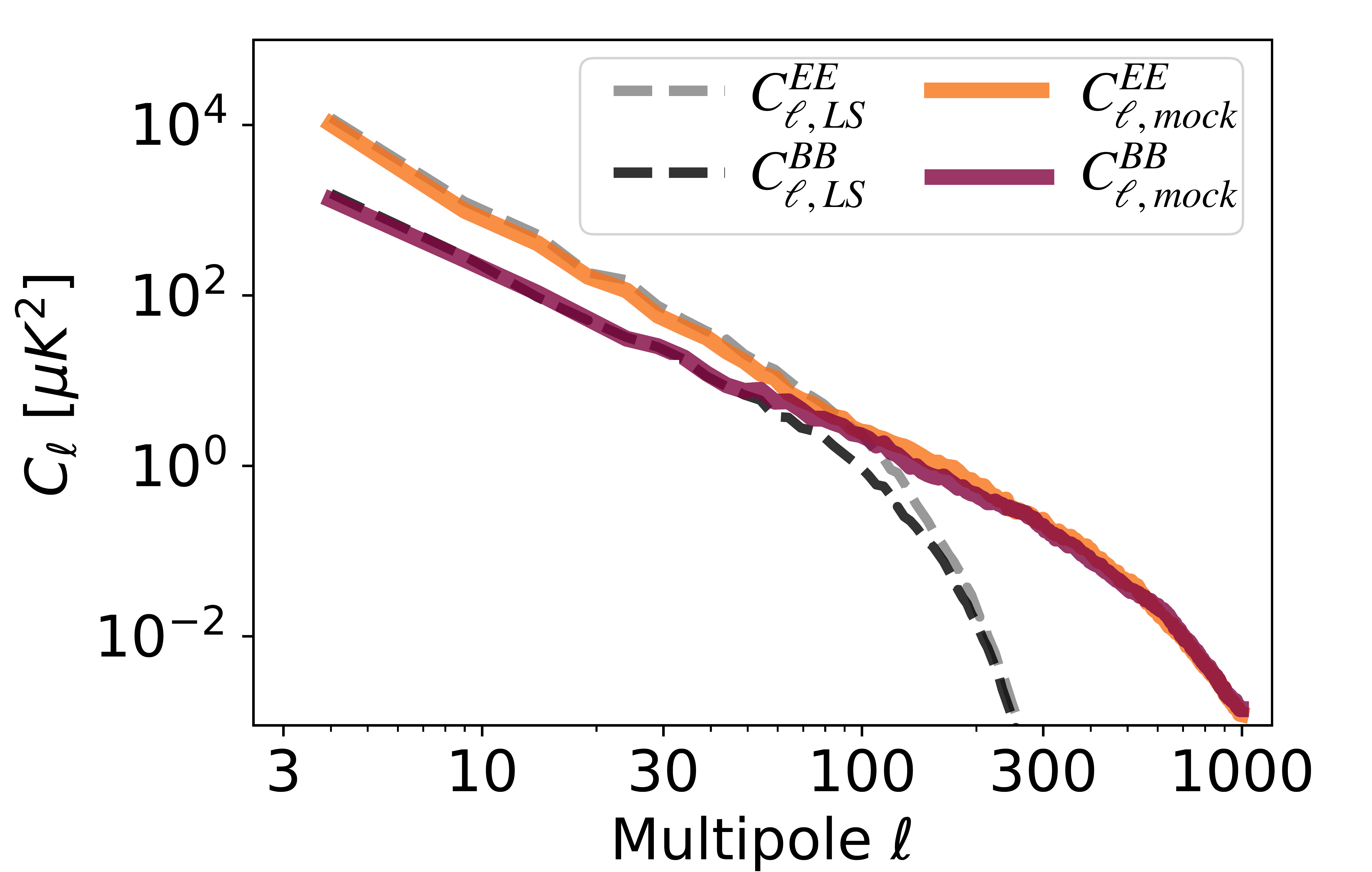}
\caption{\label{fig:spectra_fullsky} Polarization power spectra of the low resolution GNILC maps at 80 arc-minutes (dashed lines) and the final maps obtained with \forse at 12 arc-minutes (solid lines).} 
\end{figure}

Since we wanted to obtain polarization full sky maps of dust emission with the inclusion of GAN generated small scale structures, we needed to reproject the obtained square patches into a full sky HEALPix map. \par

The projection into the spherical domain is complicated by the fact that there is no large scale continuity in two consecutive patches between the high-resolution predictions performed by  \forsee. This is due to the fact that  when the neural network generates  the small scale features for a given patch, it is essentially \emph{blind} to the structures encoded at the edges of its nearest neighbors, and,  as a result, we observe a feature mismatch at the borders. In order to reduce the border effects, we (i) applied a  \emph{cosine} apodized 2D window function (see right panel of Fig.\ref{fig:reproj_scheme}) to the square patches before performing the full-sky projection  and (ii) we averaged the patches in the overlapping area with weights given by the apodization window. The details on how the projection is performed  are outlined in Appendix \ref{sec:appendix}.\par

In  Figure \ref{fig:QU_fullsky} we show the final reprojected HEALPix $Q$ and $U$ maps at angular resolution of 12 arc-minutes (with $N_{side}=2048$) compared with the input GNILC ones at 80 arc-minutes. We notice that the injection of small scales do not globally change the large scale features and, by looking at the difference between maps (third row in Figure \ref{fig:QU_fullsky}), we see that no border effects are visible. However, the application of an apodization window function and the averaging in the reprojection procedure could causes an extra-smoothing of features especially where the patch overlap is large (i.e. close to the Galactic poles). To quantify this effect we proceeded as follows: we resampled back the obtained full sky maps into square patches and computed the ratio ($r^p_{\ell}$) between the polarization power spectra of each patch before and after the reprojection. Any smoothing effect should therefore be seen as a ratio  $r^p_{\ell}>1$. Since the effect depends on the overlap of adjacent patches, which is a function of the latitude at which the tiles are centered (see left panel of Figure \ref{fig:reproj_scheme}), we averaged $r^p_{\ell}$ over all the patches located at the absolute Galactic latitude. Results are reported in Figure \ref{fig:reprj_eff} for the ratio between $EE$ spectra (similar results are obtained for $BB$).  As it can be seen the loss of power is limited, below a factor 2, at all multipoles and for all the patches centered at Galactic latitudes $| b_c |\leq72^{\circ}$ while, as expected, is more critical at the Galactic poles. \par

Finally, Figure \ref{fig:spectra_fullsky} shows the polarization power spectra computed from the final full sky $Q$ and $U$ maps compared with the ones of the input large scale maps. As expected the NN overall effect into the EE and BB power spectra results into a further extrapolation of the spectra up to $\ell\sim1000$, where we can notice the drop in power due to the resolution scale. No additional loss of power due to the reprojection is visible at the full sky level.  
We further compared the power spectra  of the input low resolution $Q$ and $U$ maps with the ones estimated from the GAN maps by estimating the index of the power law as $C_{\ell} \propto \ell ^{-\alpha} $ and we performed the fit on a different multipole range, i.e. $\ell\leq 100$ for the former and $\ell\leq 800$ for the latter. The spectral indices estimates from the large scale maps are $\alpha _{EE} =  2.48 $ and $\alpha _{BB} = 2.46 $, whereas for the NN maps we get 
  $  2.54$ and $ 2.38 $ respectively for EE and BB; indicating that the NN  does not induce any pivoting scale in the polarization power spectra at sub-degree angular scales.
  
 However,  The $E/B$ asymmetry holds up only to the scales where the dust polarization has been probed by latest Planck measurements. As already stressed, this is somewhat expected since no polarization small scale data, that encode this characteristic,  are given as training features to the GAN.  Once high resolution data will be made publicly available from the current and future high frequency experiments, even in partial regions of the sky,  we plan to update the training weights with those data sets  to improve fidelity in the generation of dust polarization features.

\section{Discussion and conclusions}
\label{sec:conclusions}

In this work we have presented a novel approach, based on the use of GANs, to simulate realistic non-Gaussian sub-degree foreground emission. This represents an important task, as in current foreground simulations small scale structures are usually injected as Gaussian random realizations of the expected power spectra. This simplification of the statistical properties of foregrounds makes it difficult to estimate their impact, especially on gravitational lensing reconstruction and delensing. Having the possibility to rely on realistic foreground simulations at all the angular scales would thus be extremely important in the preparation of the next generation of CMB experiments. To our knowledge, this is the first time that GAN methodology is used to generate polarization maps of astrophysical emission and to extend it  to smaller angular scales.\par 

We applied our algorithm, named \forsee, to thermal dust radiation considering both total intensity and polarized signal. In all the cases, we trained the GAN to generate small scale structures, at 12 arc-minutes, starting from large ones, at 80 arc-minutes. We applied the method to square sky patches with physical dimension of $20^{\circ}\times20^{\circ}$ and resolution of $320\times320$ pixels.\par
In our first test in total intensity, we could make use of available observations from the Planck satellite (at 353 GHz) of the small scale structure of thermal dust in regions close to the Galactic plane. We therefore trained the network to produce sub-degree features that mimic the statistics of the real ones. Results (shown in Figures \ref{fig:intensity_inout}, \ref{fig:intensity_minko} and \ref{fig:intensity_spectra}) demonstrate how \forse is able to achieve this goal. In particular, we characterized the obtained structures by mean of the three Minkowski functionals ($\mathcal{V}_0$, $\mathcal{V}_1$, $\mathcal{V}_2$) which are sensitive to the presence of non-Gaussianity. The agreement between the distributions computed from real and mock structures is remarkably good (see Figure \ref{fig:intensity_minko}). We  also computed the angular power spectra of the final images, showing how the amplitude of the generated features follow the correct scaling as a function of the angular scale.\par

We applied our approach also to polarized emission. Since in polarization there are not publicly available high resolution data, we made the assumption that small scale features on Stokes $Q$ and $U$ maps follow the same statistics as the one in total intensity. We trained the networks accordingly, having as target distribution the one of small scale $I$ structures. The results for polarization maps are reported in Figures \ref{fig:QU_inout}, \ref{fig:pol_minko} and \ref{fig:pol_patches}. Again, we find a very good agreement between the statistical properties of the generated structures on $Q$ and $U$ maps and the target total intensity ones, with distributions of the Minkowski functionals significantly different from those of a Gaussian field. Moreover, polarization power spectra show the correct scaling as a function of multipoles.\par

We reprojected the flat patches into the Celestial sphere, obtaining Stokes $Q$ and $U$ full sky maps of the thermal dust emission at sub-degree angular resolution which are shown in Figure \ref{fig:QU_fullsky}.
 Maps are made publicly available online\footnote{Maps can be downloaded from \url{https://portal.nersc.gov/project/sobs/users/ForSE/} }, they are stored  as a fits file and we follow the same conventions  as in the GNILC dust map  (i.e. $\mu K_{CMB}$, Galactic coordinates, same pixelization). \par
 
 The reprojection on the sphere causes an extra smoothing of the generated feature, which is particularly important in the regions close to the Galactic poles (Fig. \ref{fig:reprj_eff}).  Furthermore, due the fact that our GAN could not been trained with polarization data, the EE and BB power spectra of full sky maps reproduce the scaling observed at large angular scales but they do not preserve the $E/B$ asymmetry that have been observed at low multipole at smaller angular scales. \par
 
As in this paper we presented the method and its first application to thermal dust emission, we plan to study  the impact of the predicted non-Gaussian foreground on lensing reconstruction, de-lensing and primordial $B$-mode detection in an upcoming work. We also foresee further developments of the approach presented in this work, with several  possible applications. In particular, it could be used to further extend the angular resolution of foreground models at even smaller scales. This could be done in different ways. One is obviously by making use of new data that will become publicly available in the coming years, which will include also the observations of small scale structures in polarization, allowing to overcome the assumptions we have made in our work. The second way would be to train and apply the GAN predictions iteratively, in a sort of ``fractal''  approach, by considering the foreground statistical properties as scale invariant. One could further think of training a NN  to learn and reproduce the relation between large and small structures as well as their statistical properties obtained from  MHD simulations  and then apply it to   low-resolution data. Finally, although  we have applied \forse specifically  to thermal dust emission, as one of the strongest contaminant to CMB observations, nothing prevents to adopt it on other kind of emissions, as long as a sufficient amount of data exists. \par
 
\begin{acknowledgements}
NK acknowledges support from the ASI-Cosmos and ASI-LiteBIRD programs, from the INDARK and LiteBIRD INFN initiatives and from the ``Leonardo da Vinci'' grant (MIUR-2019). NK also thanks the Simons Foundation for hosting her at the Flatiron Institute (New York, USA) in fall 2019, and the whole group of Cosmology x Data Science at CCA (Center for Computational Astrophysics) for the extremely useful discussions at the beginning of this work. The authors thank Prof. Carlo Baccigalupi for valuable comments. This research used resources of the National Energy Research Scientific Computing Center (NERSC), a U.S. Department of Energy Office of Science User Facility operated under Contract No. DE-AC02-05CH11231. Some of the results in this paper have been derived using the HEALPix \citep{2005ApJ...622..759G} package.

\end{acknowledgements}

\bibliography{forse}{}
\bibliographystyle{aasjournal}



\appendix
\section{From Healpix to flat patches and back}
\label{sec:appendix}

In order to build the training datasets we needed to extract square images from the GNILC maps, which are projected on the sphere considering the HEALPix tessellation scheme \citep{2005ApJ...622..759G}. In particular, the flat images correspond to patches with angular dimension of $20^{\circ}\times 20^{\circ}$ in the sky and resolution of  $3.75$ arc-minutes ($320\times320$ pixels).  We chose the pixel dimension in the flat reprojection to be larger than the HEALPix one (which is $1.72$ arc-minutes for a pixelization at $N_{side}=2048$); in this way, to each pixel in the flat domain corresponds about 2 pixels in the HEALPix one and we further reduced the possibility of having missing pixels in the flat projection.  \par

To perform the projections we made use of \texttt{healpy} \citep{Zonca2019}, \texttt{astropy}\footnote{\url{https://www.astropy.org/}  } and the \texttt{reproject}\footnote{\url{https://reproject.readthedocs.io/en/stable/index.html}.} python packages. 

\begin{figure}[b]
\centering
\includegraphics[width=11cm]{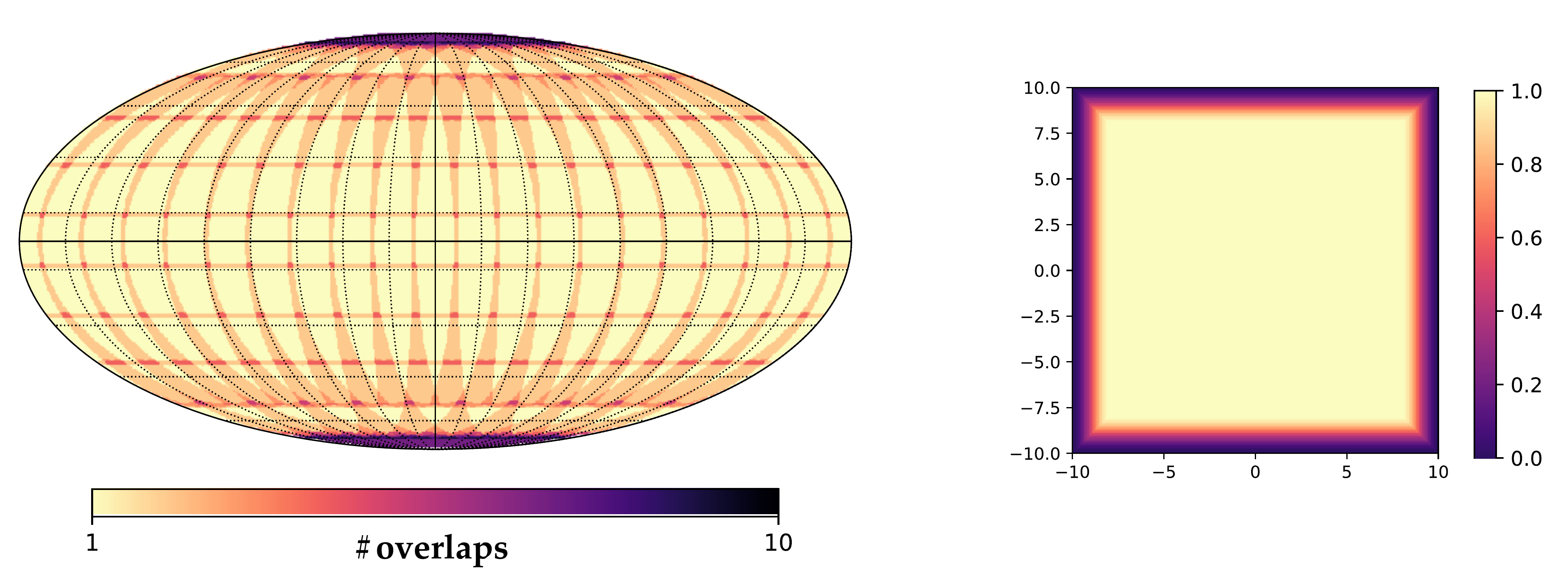}
\caption{Left panel: map of overlaps given the scheme used to make the projections. Right panel: apodization window adopted to project the square patches into Celestial sphere with the HEALPix gridding.}\label{fig:reproj_scheme}
\end{figure}

\begin{description}
\item [From HEALPix to flat] for the polarization case, we needed to get flat images that cover the whole Celestial sphere, in order to generate small scale features everywhere in the sky. In particular, we considered 174 locations, corresponding to the center of  our $20^{\circ}\times 20^{\circ}$ patches.
The patch centers were taken from a grid, with number of patches in Galactic longitude changing as a function of Galactic latitude. In particular, the grid encodes 10 locations in Galactic latitude equally spaced by $18^{\circ}$, with each patch overlapping by $2^{\circ}$. The size of the grid adaptively changes in Galactic longitude: namely 7 patches are centered at $b=\pm 90^{\circ}$ and oriented at equally spaced longitudes, 10  patches at $b=\pm 72^{\circ}$  and 20 patches for the other latitude centroids. This choice was motivated by finding a trade off between having a full and  homogeneous coverage of the sphere and the necessity to include some degree of overlap between patches in order to minimize border effects (see Section \ref{sec:fullsky}).   

In the left panel of Figure \ref{fig:reproj_scheme} we show for each pixel in the HEALPix map the number of flat patches that includes it. The degree of overlap increases with the Galactic latitude, reaching its maximum at the Galactic Poles.
 
\item [From flat to HEALPix] as described in Section \ref{sec:fullsky} we needed to apply an apodization to the small scale patches generated by \forse before reprojecting them into the sphere, in order to reduce the border effects. In particular, we applied a \emph{cosine} apodized 2D window function, $W_{apo}$, defined as : 
\begin{displaymath}
W_{apo} (x)= \left\{  
		\begin{array}{ll}
			\frac{1}{2}\left(  1 - \cos( \pi x  )  \right)  & \mbox{if } x < 1  \\
		1 & \mbox{otherwise }  
	\end{array}
	\right. 
\end{displaymath}
where $x\equiv \sqrt{(1-\cos \eta)/(1 - \cos \eta_{apo} )}$, $\eta$ being the angular separation between the pixels and the edge of the image, $\eta_{apo}$ the apodization length chosen to be the same as the overlap  scale (i.e. $2^{\circ}$). The resulting apodization window is shown in the right panel of Figure  \ref{fig:reproj_scheme}. The patches were then combined in the overlapping area as a weighted average with weights given by  $W_{apo}$ into a single HEALPix map. 
\end{description}

\end{document}